# Differences in the magnon diffusion length for electrically and thermally driven magnon currents in $Y_3Fe_5O_{12}$

Juan M. Gomez-Perez,[1] Saül Vélez,[1,2,*], Luis E. Hueso,[1,3] Fèlix Casanova[1,3,*]

[1]CIC nanoGUNE BRTA, 20018 Donostia-San Sebastian, Basque Country, Spain
[2]Department of Materials, ETH Zürich, 8093 Zürich, Switzerland
[3]IKERBASQUE, Basque Foundation for Science, 48013 Bilbao, Basque Country, Spain

[*]Email: saul.velez@mat.ethz.ch; f.casanova@nanogune.eu

**Abstract**

Recent demonstration of efficient transport and manipulation of spin information by magnon currents have opened exciting prospects for processing information in devices. Magnon currents can be driven both electrically and thermally, even in magnetic insulators, by applying charge currents in an adjacent metal layer. Earlier reports in thin yttrium iron garnet (YIG) films suggested that the diffusion length of magnons is independent on the biasing method, but different values were obtained in thicker films. Here, we study the magnon diffusion length for electrically and thermally driven magnon currents in the linear regime in a 2-μm-thick YIG film as a function of temperature and magnetic field. Our results show a decrease of the magnon diffusion length with magnetic field for both biasing methods and at all temperatures from 5 to 300 K, indicating that sub-thermal magnons dominate the long-range transport. Moreover, we demonstrate that the value of the magnon diffusion length depends on the driving mechanism, suggesting that different non-equilibrium magnon distributions are biased for each method. Finally, we demonstrate that the magnon diffusion length for thermally driven magnon currents is independent of the YIG thickness and material growth conditions, confirming that this quantity is an intrinsic parameter of YIG.

## I. Introduction

Insulator-based spintronics is nowadays one of the most promising fields to transport and manipulate spin information, which in magnetic insulators (MIs) is carried by magnons [1–21] instead of conduction electrons in non-magnetic conductors [22]. A magnon is a quasiparticle corresponding to a collective magnetic-moment precession that carries a spin angular momentum in materials with magnetic ordering, i.e., a quantized spin wave. The possibility of transporting spin information over long distances using magnons [3,8,16], combined with the fact that the use of MIs prevents both the emergence of spurious transport effects due to charge flow in the ferromagnet and unnecessary Ohmic losses, make MIs very attractive materials for spintronic applications with potential to compete with their metallic counterparts.

Magnon spin currents in MIs can be driven either by ferromagnetic resonance [8,9] (low-frequency coherent magnons), by a thermal gradient [16] (high-frequency incoherent magnons), or by electrical means by making use of the spin Hall effect (SHE) of heavy metals such as Pt [3] (high-frequency incoherent magnons), whereas, in most of the cases, have been detected electrically by employing the inverse spin Hall effect (ISHE). In the original work of Cornelissen *et. al.* [3], magnon currents were electrically and thermally driven in the magnetic insulator yttrium iron garnet $Y_3Fe_5O_{12}$ (YIG), and electrically detected up to tens of micrometers apart at room temperature, by employing a nonlocal configuration in which Pt strips acted as injectors and detectors of the magnon currents. Many studies on incoherent magnon spin transport have followed [3–6,10,12,13,15–19,21], in which YIG played the central role because of its soft ferrimagnetism, negligible magnetic anisotropy, and low Gilbert damping [3–6,8,12,16,18,19,21,23], albeit magnon spin transport has also been demonstrated in other MIs [11,13–15,17].

The key parameter that defines the characteristic length to which magnons propagate is the magnon diffusion length ($\lambda_m$). Its value has been studied in YIG as a function of temperature [4,5,10], magnetic field [18], thermal gradient [16], and thickness [21]. Earlier reports suggested that $\lambda_m$ is the same regardless of the mechanism by which the magnons are biased (by thermal gradients or by torques employing the SHE, in the linear-response regime for magnon spin and heat transport) [4], but more recent works showed that those values might be different [10,21]. It is thus not clear whether the thickness of the MI layer might play a role on the value of $\lambda_m$ or whether the distribution of non-equilibrium magnons that contribute to the magnon transport are different due to the different nature of the biasing methods employed.

In this paper, by employing a nonlocal configuration, we report a systematic temperature and magnetic field dependence study of the magnon diffusion length in 2-μm-thick YIG films for electrically and thermally driven magnon currents in the linear-regime. Our results evidence that the diffusive transport of the magnon currents depends on the way they are driven, supporting the idea that the two methods bias different non-equilibrium magnon distributions. Furthermore, we demonstrate that the magnon diffusion length of the thermally driven magnon currents, and its temperature dependence, are the same regardless of the YIG thickness and growth method. This result shows the robustness of thermally induced magnon transport to extract the magnon transport properties of YIG and, by extension, of MIs.

**II. Experimental details**

The devices were fabricated on top of 2-μm-thick YIG films provided by Innovent e.V. (Jena, Germany). YIG was grown by liquid phase epitaxy (LPE) in a (111) gadolinium gallium garnet ($Gd_3Ga_5O_{12}$, GGG) substrate. In a first sample (sample 1), a 5-nm-thick Pt layer was magnetron-sputtered *ex situ* (80 W; 3 mtorr of Ar) on top of YIG, and Pt strips (width $w = 450$ nm, length $L = 80$ μm) were patterned by negative e-beam lithography and Ar-ion milling with different edge-to-edge distances ($d = 1-20$ μm). In a second sample (sample 2), devices with distances ranging from 8.5 μm to 125 μm were fabricated. In this case, for shorter distance devices ($d = 8.5-40$ μm), the Pt strip dimensions are the same as for sample 1, whereas for longer distances ($d = 50-125$ μm), the dimensions are $L = 650$ μm and $w = 2.7$ μm and the measured voltage is normalized accordingly [3]. A scanning electron microscopy (SEM) image of one of the devices is shown in Fig. 1(a).

Magnetotransport measurements were carried out in a liquid-He cryostat at temperatures $T$ between 10 and 300 K, externally applied magnetic fields $H$ up to 9 T, and a 360º in-plane sample rotation. We use a nonlocal configuration, where we apply a DC current $I$ along a Pt strip (injector) while we measure the voltage along an adjacent Pt strip (detector). In order to separate the electrically generated nonlocal voltage from the thermally generated one, the polarity of the DC current is reversed at the injector. The difference of the nonlocal voltage, $V_{NL}^e = [V_{NL}(+I) - V_{NL}(-I)]/2$, corresponds to the electrical signal (equivalent to the 1$^{st}$ harmonic signal when using an AC measurement), whereas the sum, $V_{NL}^{th} = [V_{NL}(+I) + V_{NL}(-I)]/2$, corresponds to the thermally driven signal (equivalent to the 2$^{nd}$ harmonic signal) [6,24,25].

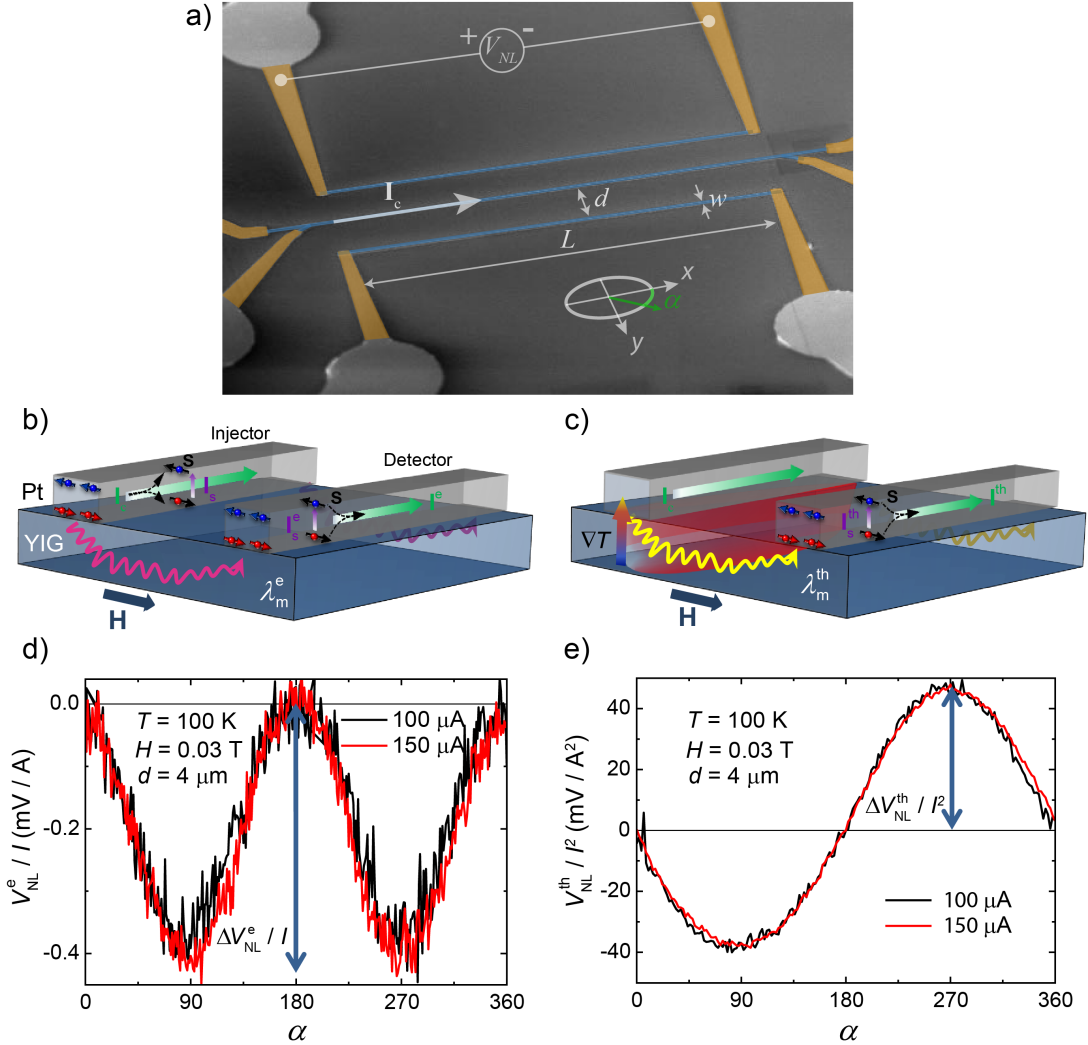

FIG. 1. (a) SEM image of a nonlocal device used in this work ($d$ = 6 μm). The measurement configuration and the geometrical parameters are indicated. (b),(c) Schematic representation of the creation, transport and detection of magnon currents in the nonlocal devices by means of (b) electrical biasing and (c) thermal biasing. $\mathbf{I}_s$ indicates the spin current generated by $\mathbf{I}$, $\mathbf{s}$ the vector spin, $\mathbf{I}_s^e$ ($\mathbf{I}_s^{th}$) the spin current generated at the detector for electrically (thermally) driven magnon currents, $\mathbf{I}^e$ ($\mathbf{I}^{th}$) the charge current generated at the detector due to the spin to change conversion for the electrically (thermally) driven case, $\lambda_m^e$ ($\lambda_m^{th}$) the magnon diffusion length for electrically (thermally) driven magnons, and $\nabla\mathbf{T}$ indicates the thermal gradient. (d),(e) Representative angular-dependent nonlocal signals detected for (d) electrically and (e) thermally driven magnon currents using the measurement configuration shown in (a). The magnetic field applied is $H$ = 0.03 T, which is sufficient to saturate the magnetization of our YIG films. The measurement temperature is 100 K and the device corresponds to sample 1. The nonlocal signals are presented as the nonlocal voltage normalized to the applied current in (d) and to the square of the applied current in (e), showing that $V_{NL}^e$ and $V_{NL}^{th}$ are linear and quadratically dependent on the injected current, respectively. The amplitude of the signals, $\Delta V_{NL}^e/I$ and $\Delta V_{NL}^{th}/I^2$, are indicated with solid arrows.

### III. Results and Discussion

**A. Angular dependence of the nonlocal signal for electrically and thermally driven magnon currents.**

We investigate the magnon spin transport for both electrically and thermally induced magnon currents in YIG by employing the nonlocal configuration shown in Fig. 1(a). This device scheme allows exploiting the large SHE in Pt [26–29] to electrically generate and detect magnon currents in YIG. By applying a charge current along a Pt strip, a spin accumulation is induced at the Pt side of the Pt/YIG interface due to the SHE [which is in the plane of the film and perpendicular to the current, see vector notation in Fig. 1(b)]. When the YIG magnetization and spin polarization of the spin accumulation in Pt are parallel (antiparallel), a magnon is annihilated (created) due to exchange interaction between the Pt electron spins and the YIG magnetic moments, leading to a magnon imbalance that modifies the magnon chemical potential close to the interface [2]. This gives rise to a diffusion of magnons (magnon spin current) that can propagate for several microns along YIG [2,3]. By the reciprocal process, a second Pt strip can detect the magnon imbalance, as the induced spin accumulation in Pt (due to the magnon-to-spin conversion at the interface) is finally converted to a voltage by the ISHE [9]. For weak and moderate excitation amplitudes, the electrical generation of magnon currents is a linear-response process [2,3], meaning that the detected nonlocal voltage $V_{NL}^e$ is proportional to the applied charge current (see section II for details). This is indeed confirmed in our devices for all experimental conditions reported in this work. As an example, we show in Fig. 1(d) the nonlocal $V_{NL}^e$ signal measured in a representative device at two different currents while rotating the in-plane magnetic field. The two curves nicely overlap once the signal is normalized by the injected current. Moreover, note that the angular dependence of the nonlocal voltage shows the expected *sin²* dependence, which is due to the symmetry of the SHE and the ISHE at the injector and detector Pt strips, respectively, and their relative orientation with the magnetization of the YIG layer [3,6,19]. From this plot, the amplitude of the electrically induced signal $\Delta V_{NL}^e/I$ is obtained.

On the other hand, magnon currents can also be thermally induced. In ferromagnetic materials, a thermal gradient drives a magnon spin current parallel to the induced heat flow due to the spin Seebeck effect [30]. Therefore, by making use of the Joule dissipation in a Pt strip, a thermal gradient can be generated in the YIG film beneath, resulting in a diffusive magnon current that can be nonlocally detected by employing the ISHE of a second Pt strip (Fig. 1c) [3–5,10,12,16,21]. In this nonlocal spin Seebeck configuration, and in the linear-response regime of thermally driven magnon currents, the voltage $V_{NL}^{th}$ scales quadratically with the applied charge current [2,3] (see section II for details). This is confirmed in our devices at temperatures from 5 to 300 K and for the whole range of currents employed in this work. This dependence is demonstrated in Fig. 1(e), where we show the angular dependence of the nonlocal voltage $V_{NL}^{th}$ measured at two different currents in a representative device, which coincide once we normalize the curves to the square of the injected current. At 2.5 K, however, $V_{NL}^{th}$ does not scale with the square of the current (see appendix A), an anomaly already reported in recent works [31,32]. For this reason, the magnon diffusion length is not evaluated at this temperature. The angular dependence of the thermally driven voltage shows a *sin* dependence [see Figs. 1(a) and 1(e)], because in this case only the ISHE symmetry of the detector plays a role [3,6]. From this plot, the amplitude of the thermally induced signal $\Delta V_{NL}^{th}/I^2$ is obtained.

From these two types of nonlocal signals, and by measuring their amplitude in devices having different distances *d* between the injector and detector strips, we can extract the magnon spin diffusion length of our YIG films.

**B. Temperature and magnetic-field dependence of the nonlocal signals.**

We now investigate the temperature- and magnetic field-dependence of the magnon transport for both thermally and electrically driven magnon currents. The temperature dependence of the amplitude of the nonlocal signal, measured at $H = 0.03$ T and for different distances between the Pt electrodes, is plotted in Figs. 2(a) (for electrically biased magnons) and 2(b) (for thermally biased magnons).

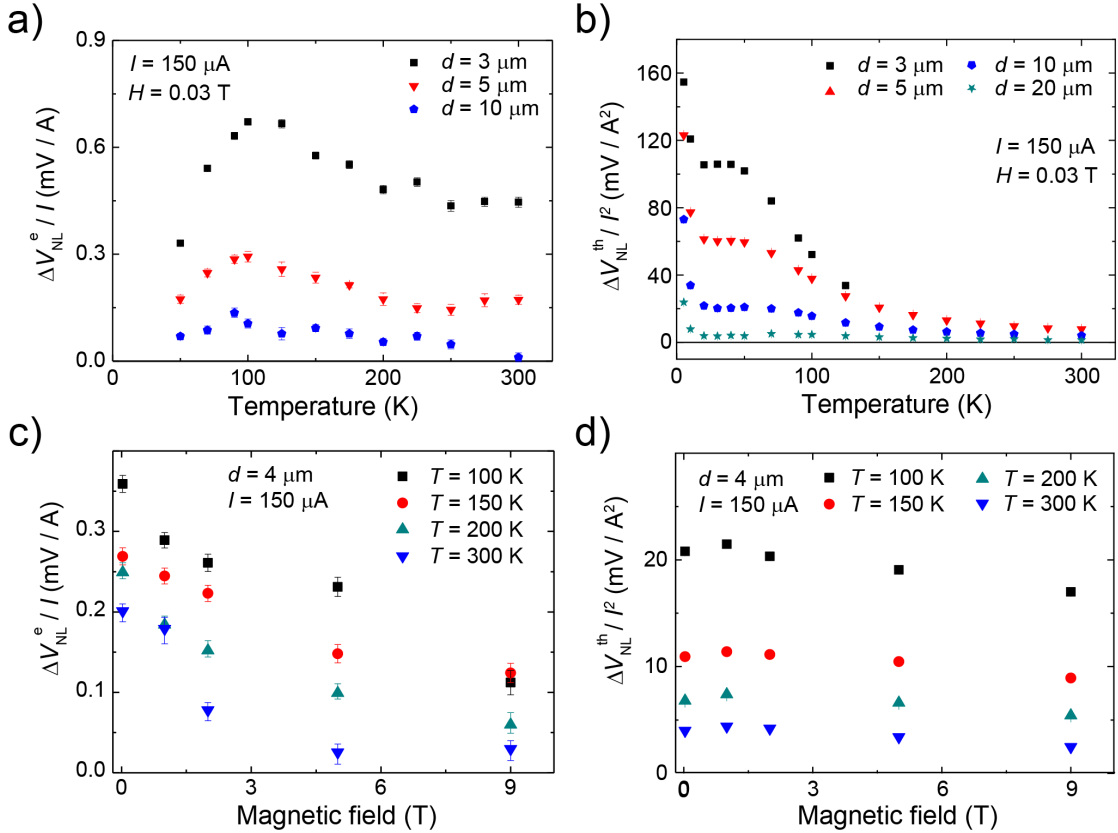

FIG. 2. (a-b) Temperature dependence of the nonlocal voltage amplitude for (a) electrically and (b) thermally biased magnons at 0.03 T, and for different distances between the injector and detector Pt strips. (c-d) Magnetic field dependence of the nonlocal voltage amplitude measured at a distance of 4 μm and different temperatures for (c) electrically and (d) thermally driven magnon currents. The nonlocal signal is normalized to the charge current applied (150 μA) in (a) and (c), and to the square of the current in (b) and (d). All data correspond to sample 1.

The amplitude of the nonlocal signal for electrically biased magnons $\Delta V_{NL}^e/I$ vanishes when approaching zero temperature [see Fig. 2(a)]. For the longest distances ($d \geq 10$ μm), amplitude can only be extracted reliably above 50 K and, for clarity, data below this temperature is not included for any *d*. This behavior can be understood by taking into account that the magnon population is strictly zero at zero temperature and increases with temperature, leading to a more efficient generation as the temperature raises [4,6]. For distances $d \geq 2$ μm [see Fig. 2(a) for representative measurements], the signal reaches a maximum at approximately 100 K, from which it monotonously decreases up to the highest temperature explored, 300 K. For shorter distances ($d \leq 1$ μm; not shown), the signal monotonously increases up to 300 K, which is consistent with previous results reported for very short injector-detector distances [6,19]. Besides, as expected, the nonlocal signal decreases when the distance between the Pt electrodes increases.

In the case of thermally biased magnons [Fig. 2 (b)], the largest nonlocal spin Seebeck amplitude $\Delta V_{\text{NL}}^{\text{th}}/I^2$ is found at the lowest temperature (5 K). The signal-to-noise ratio for thermally driven magnon currents is larger than for electrically driven ones, allowing us to detect nonlocal thermal signals for longer distances (~20 μm in sample 1, and up to 125 μm in sample 2). The amplitude sharply decreases between 5 K and 10 K, followed by a plateau between 10 and 50 K, and finally decreases monotonically with increasing temperature. The low temperature behavior (T < 10 K) could be related to the strong variation of the thermal gradient generated at the Pt injector, since many of the parameters defining the thermal gradient (such as the thermal conductivity and the specific heat of YIG) have a strong temperature dependence at this low temperature regime [33]. Indeed, a puzzling behavior of the nonlocal spin Seebeck effect signal have been reported at this temperature range, whose origin remains unclear [4,10]. Moreover, and in agreement with what is observed for electrically-driven magnon currents, the signal gets smaller when increasing the injector-detector distance. Finally, a sign change of the nonlocal spin Seebeck effect is expected for injector-detector distances that are below certain value (which is on the order of the YIG thickness) [12,20,21]. In our case, however, we do not observe this behavior as all measurements are taken above the distance at which the sign change occurs.

The magnetic field dependence of the nonlocal signal is plotted in Figs. 2(c) (for electrically biased magnons) and 2(d) (for thermally biased magnons) for a distance $d = 4$ μm between the Pt electrodes and different temperatures. The nonlocal signal for electrically biased magnons [Fig. 2(c)] decreases monotonously with magnetic field, being strongly reduced at 9 T. In contrast, the signal for thermally biased magnons [Fig. 2(d)] shows a shallow maximum at ~1 T, and slowly decreases to 9 T. These behaviors are similar to the ones reported in Ref. [18].

**C. Temperature and magnetic field dependence of the magnon diffusion length.**

From the data reported in the previous section, we can now extract the magnon diffusion length at different magnetic fields and temperatures, which is the relevant parameter that describes the magnon transport behavior. In order to justify the use of data points acquired in different devices (with different injector-detector distances $d$), we require that the YIG/Pt interface is the same for all devices used. We confirm this by measuring the spin Hall magnetoresistance in the different Pt strips, which show similar amplitude (see Appendix B). As an example, we plot in Figs. 3(a) and 3(b) the nonlocal signals as a function of the distance $d$ between the Pt electrodes for a particular temperature (150 K) and selected magnetic fields (0.03, 5 and 9 T). We identify two distinctive regions known as the diffusion regime (light purple region) and the exponential regime (light brown region). In the diffusion regime, which corresponds to $d \lesssim \lambda_{\text{m}}$, the non-equilibrium magnon accumulation diffuses away from the injector into the YIG with a geometrical decay [2,3,10]. Further away, at $d \gtrsim \lambda_{\text{m}}$, magnons relax showing an exponential decay for the two type of biased magnon currents [10]. Therefore, in order to determine $\lambda_{\text{m}}$, a linear fit of the natural logarithm to the nonlocal voltage amplitude is performed in the exponential regime, as shown in Figs. 3(a) and 3(b) (red solid lines). In the case of the thermally biased magnons, the linear fits should be carried out for distances shorter than 3–5 times $\lambda_m$ to avoid entering the $1/d^2$–regime, where the signal is dominated by the temperature gradient induced (by geometric thermal diffusion) at the YIG/GGG interface underneath the Pt detector [10]. From these fittings, we extracted the magnon diffusion length for both electrically ($\lambda_m^e$) and thermally ($\lambda_m^{th}$) biased magnon currents at different temperatures and applied magnetic fields. The results are presented in Figs. 3(c)-3(f). See Appendix C for a further discussion on the use of the one- or the two-dimensional diffusion model to extract the magnon diffusion length.

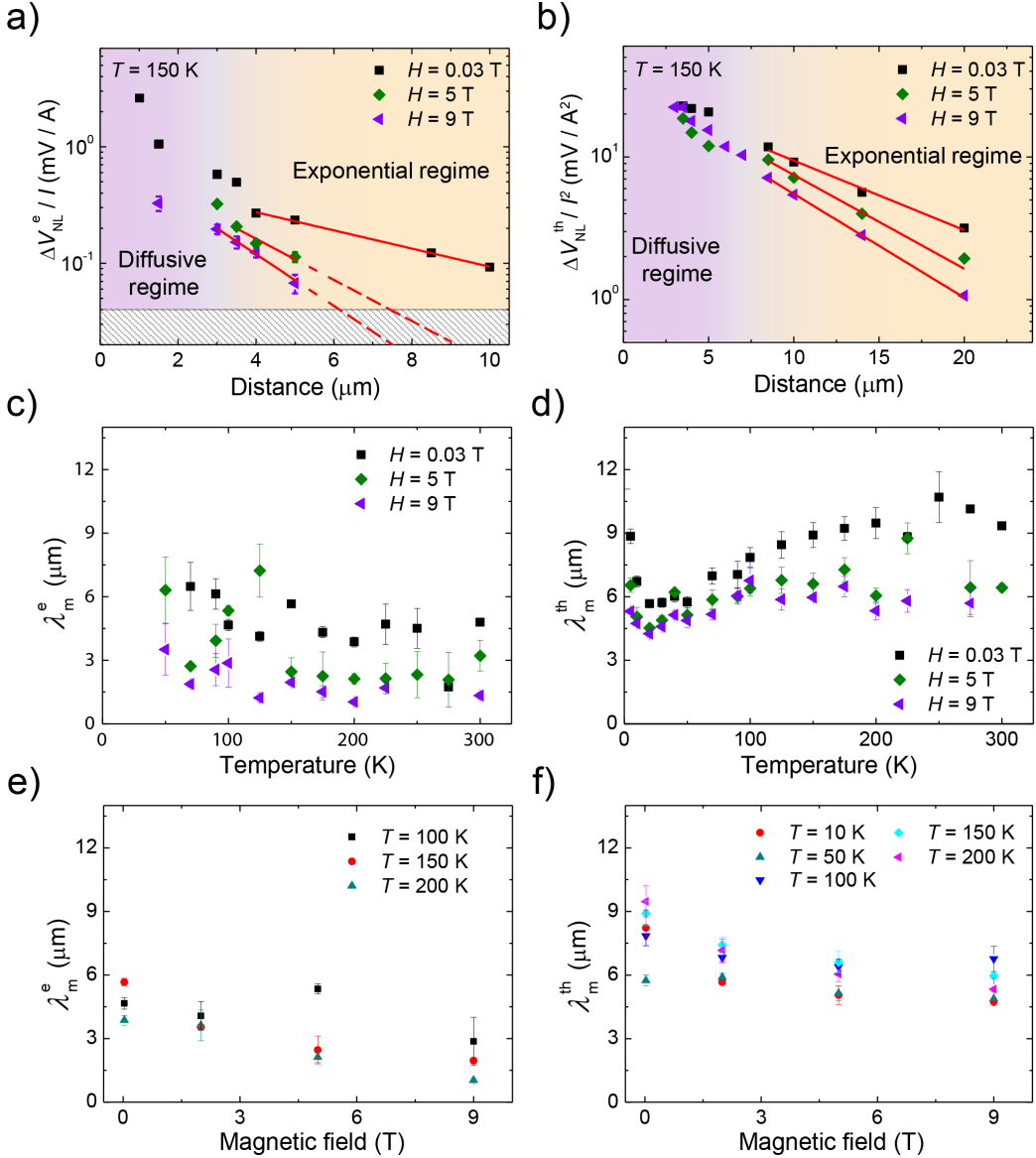

FIG. 3. (a-b) Nonlocal voltage amplitude for (a) electrically and (b) thermally driven magnon currents as a function of injector-detector distance and for different magnetic fields. The diffusive and exponential magnon transport regimes, which are indicated with different colors, are identified from the different decay of the signal with increasing distance. The red lines are fits in the exponential regime to extract the magnon diffusion length at the corresponding magnetic field applied. The dashed region at the bottom of (a) indicates the noise threshold of the measurement setup. (c-d) Temperature dependence of the magnon diffusion length for (c) electrically ($\lambda_m^e$) and (d) thermally ($\lambda_m^{th}$) driven magnon currents at different magnetic fields. (e-f) Magnetic field dependence of (e) $\lambda_m^e$ and (f) $\lambda_m^{th}$ at different temperatures. All data correspond to sample 1.

At any magnetic field applied, $\lambda_m^e$ shows a constant or slightly decreasing behavior with increasing temperature [Fig. 3(c)]. As discussed above, we cannot estimate $\lambda_m^e$ below 50 K because of the small signal-to-noise ratio. $\lambda_m^e$ decreases monotonously with increasing the magnetic field [Fig. 3(e)], in line with the strong decay reported in Ref. [18] for electrically biased magnon currents at 300 K. Both the temperature and field dependences are completely different for $\lambda_m^{th}$ [Fig. 3(d)]. At low magnetic fields (0.03 T), $\lambda_m^{th}$ is maximum at the lowest measured temperature (5 K) and decreases up to 30 K. In this temperature range, the increase of the magnetic field reduces the value of $\lambda_m^{th}$, although the same trend with temperature is kept in the whole range of magnetic fields explored (up to 9 T). For temperatures above 30 K, however, two different temperature dependences for $\lambda_m^{th}$ are identified depending on the strength of the magnetic field: i) at low magnetic fields (0.03 T), $\lambda_m^{th}$ slightly increases with temperature, in contrast with the behavior of $\lambda_m^e$ for the same magnetic field and temperature range; ii) at high magnetic field (5 T and 9 T), $\lambda_m^{th}$ becomes fairly temperature independent. It is also worth mentioning that the decay of $\lambda_m^{th}$ with increasing magnetic field is rather similar at all temperatures, showing a strong suppression in the low field regime ($\leq$ 2 T), from which a relatively smaller reduction proceeds up to 9 T [see Fig. 3(f)]. This behavior is in agreement with the reported field-dependence at 300 K in a thin (0.2 μm) YIG film [18]. The origin of the differences between the magnon diffusion lengths will be discussed in the following section.

**D. Comparison between electrical and thermal magnon diffusion length.**

In the following, we compare the temperature dependence of $\lambda_m^e$ and $\lambda_m^{th}$ for low (0.03 T) and high (9 T) magnetic fields applied (see Fig. 4).

*Low magnetic field regime.* Let us start with the case of a low magnetic field applied (0.03 T). As discussed before, the relatively low signal generated for electrically biased magnon currents prevented us to measure $\lambda_m^e$ at low temperatures and, consequently, to compare the two magnon diffusion lengths in this temperature range. In the ~70–100 K range, it seems that both $\lambda_m^e$ and $\lambda_m^{th}$ converge to ~6 μm. Increasing the temperature, however, results in a splitting of both characteristic lengths, with $\lambda_m^{th}$ increasing while $\lambda_m^e$ slightly decreasing, reaching ~9.3 μm and ~4.8 μm, respectively, at room temperature. This behavior is in stark contrast to previous results reported in 0.2-μm-thick YIG films, where both magnon diffusion lengths coincide [4,18]. However, as in our 2-μm-thick YIG, recent results in thicker YIG films indicate that the magnon diffusion lengths of electrically and thermally generated magnon currents might be different [10,21], suggesting that the thickness of the YIG layer might have an influence in the magnon diffusion length.

In order to rule out the possibility of nonlocal thermal effects influencing the signals, which would lead to a misestimation of $\lambda_m^{th}$, we measured a second sample (sample 2) with devices with longer injector-detector distances. As discussed in Appendix D, we can clearly distinguish the exponential regime, from which $\lambda_m^{th}$ can be extracted reliably, from the *1/d²*–regime, which is dominated by the thermal gradient generated at the YIG/GGG interface underneath the detector [10]. By fitting the data to the entire exponential regime (which ranges up to ~40 μm), we obtain $\lambda_m^{th}$ values that are in good agreement with the values extracted from sample 1 (see Fig. 5). We thus conclude that both $\lambda_m^{th}$ and $\lambda_m^e$ are properly evaluated for the case of a low magnetic field applied and, more importantly, they are different in the 100–300 K temperature range.

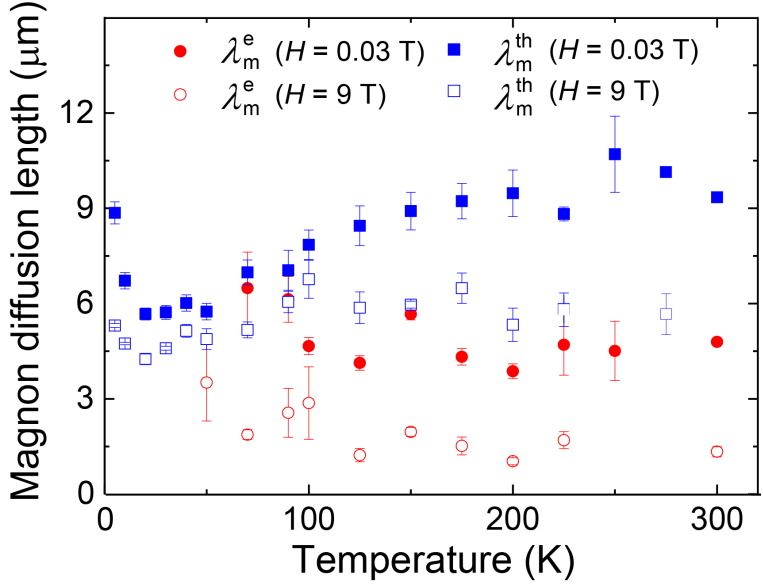

FIG. 4. Temperature dependence of the magnon diffusion length at 0.03 T (solid symbols) and 9 T (open symbols) for electrically (red circles) and thermally (blue squares) biased magnon currents. Data corresponding to sample 1.

The large difference in the magnon diffusion length observed for electrically and thermally driven magnon currents (which approaches a factor of two at room temperature) evidences that the non-equilibrium magnon distributions generated at the YIG/Pt-injector interface diffuse and relax differently depending on the way they are excited. First reports in YIG films thinner (0.2 μm) than ours (2.0 μm), however, suggested that both methods result in an equivalent biasing and diffusion of the induced non-equilibrium magnon distributions. Consequently, it was assumed that the magnon transport in the linear regime can be described by the same transport coefficients independently of the biasing method, being the long-range transport dominated by the magnon chemical potential [2]. In contrast to these earlier reports, more recent experiments in thicker films showed hints that the magnon diffusion length of electrically driven magnon currents may differ from those driven thermally [10,21], with estimated $\lambda_m^e$ values that are surprisingly similar to the one extracted here ($\lambda_m^e \sim 6$ μm at room temperature [21]). The existence of such discrepancy, however, was not discussed.

In order to understand the origin of the discrepancy between $\lambda_m^e$ and $\lambda_m^{th}$, we first need to identify the ingredients that may contribute to the magnon diffusion length. While current theories assume that thermal magnons are the ones that dominate the magnon transport [2], studies of the longitudinal spin Seebeck effect in YIG/Pt heterostructures indicate that sub-thermal magnons, i.e., low-frequency magnons with energies below $k_B T$, are the ones that dominate the long-range transport [33–36]. This interpretation is inferred from the suppression of the SSE signal with magnetic field [34,35,37], which persist at temperatures much above the field-induced Zeeman splitting (~10.7 K at 8 T [35]). Our experiments show a significant suppression of the magnon diffusion length with magnetic field at all temperatures for both biasing methods [Figs. 3(e) and 3(f)], thus suggesting sub-thermal magnons to dominate the long-range magnon transport for both thermally and electrically driven magnon currents. In fact, it has been experimentally demonstrated that the low-frequency magnons exhibit much longer spin lifetimes than the high-frequency magnons, thus dominating bulk magnon transport due to the fast cooling of the high-energy (thermal) magnons [36]. Nevertheless, the magnon diffusion length is the result of integrating in frequency the whole distribution of biased non-equilibrium magnons, taking into account their group velocity and spin lifetimes in a non-linear fashion [38]. We thus conclude that, while our results evidence that sub-thermal magnons dominate the long-range transport for both electrically and thermally driven magnon currents, the difference in the magnon diffusion length observed between the two methods might arise from the biasing of different non-equilibrium magnon distributions. We also note that, besides the magnon chemical potential, recent results in the YIG thickness dependence of the longitudinal spin Seebeck effect show that the magnon energy relaxation length (the distance at which the magnon and phonon temperatures equilibrate) also influences the diffusive spin transport [39,40]. Although this length is reported to be one order of magnitude shorter than the magnon diffusion length (~250 nm compared to ~5–10 μm), it evidences that the two biasing methods cannot be considered fully equivalent.

*High magnetic field regime.* We now turn to the case of high magnetic fields (>2 T), a regime in which both magnon diffusion lengths reduce with respect to the low field case (compare, for instance, open and solid symbols in Fig. 4). In this case, the difference between $\lambda_\mathrm{m}^\mathrm{th}$ and $\lambda_\mathrm{m}^\mathrm{e}$ increases up to a factor of four for temperatures above 100 K. This large difference is due to the strong decay of $\lambda_\mathrm{m}^\mathrm{e}$ with magnetic field [Fig. 3(e)], whereas $\lambda_\mathrm{m}^\mathrm{th}$ shows a much weaker reduction with *H* above ~2 T [Fig. 3(f)]. This different behavior of the diffusion lengths with magnetic field was already reported by Cornelissen *et al.* in a 0.2-μm-thick YIG at room temperature and was interpreted as an artifact arising from the temperature gradients present close to the detector [18].

Indeed, the influence of such thermal gradients gives rise to long-ranged nonlocal signals (*1/d²*–regime), which are expected to dominate at distances longer than 3–5 times the magnon diffusion length [11]. Although we have ruled out this possibility at low fields (Appendix D), the decay of $\lambda_\mathrm{m}^\mathrm{th}$ with increasing the magnetic field [Fig. 3(f)] might lead to a shift of the *1/d²*–regime to shorter distances. Whereas at 2 T we can still clearly distinguish the exponential regime from the *1/d²*–regime, at 5 and 9 T they cannot be distinguished (see Appendix D), suggesting us that the apparent saturation of $\lambda_m^{th}(H)$ above 2 T could arise from an overestimation of the magnon diffusion length caused by the fitting of the nonlocal signal over distances partially influenced by the *1/d²*–regime. We note, however, that a different magnetic-field dependence of $\lambda_\mathrm{m}^\mathrm{e}$ and $\lambda_\mathrm{m}^\mathrm{th}$ could also be explained by the contribution of different non-equilibrium magnon distributions to the magnon transport depending on the biasing method, in line with the discussion given above for the low field regime.

**E. Robustness of the thermal magnon diffusion length with YIG thickness.**

In the following, we show that the temperature dependence of the magnon diffusion length of the thermally driven magnon currents at low fields is the same regardless of the YIG thickness and crystal growth method. In Fig. 5, we present the temperature dependence of $\lambda_\mathrm{m}^\mathrm{th}$ for our 2-μm-thick YIG (blue solid squares for sample 1 and cyan solid circles for sample 2), together with the data of two previous works: a single crystal of YIG (500 μm) grown by the Czochralski method [5] (black open triangle), and a thinner

YIG with a thickness of 0.2 μm grown by LPE [4] (red open triangle). Interestingly, the magnon diffusion lengths in all four samples have the same temperature dependence and comparable value. This surprisingly good match evidences that the YIG thickness and the growth method are not relevant parameters that determine how far thermally driven magnon currents can flow through YIG, showing the robustness of the extracted values. Considering that these are independent measurements in three different experimental setups, but that in all of them a local thermal gradient is generated for creating the magnon currents, we conclude that $\lambda_m^{th}$ is an intrinsic parameter of YIG that is associated to the diffusion of thermally biased magnons. We should mention here that the temperature dependence of $\lambda_m^{th}$ for a 2.7-μm-thick YIG sample (a thickness similar to ours) is also studied in Ref. [10]. $\lambda_m^{th}$ has the same trend with temperature, but with larger values: $\lambda_m^{th}$ at room temperature is ~15 μm and the maximum value at the lowest temperature rises up to ~40 μm. We do not find an easy explanation for the difference between our values and the ones reported in Ref. [10], which might need further investigation.

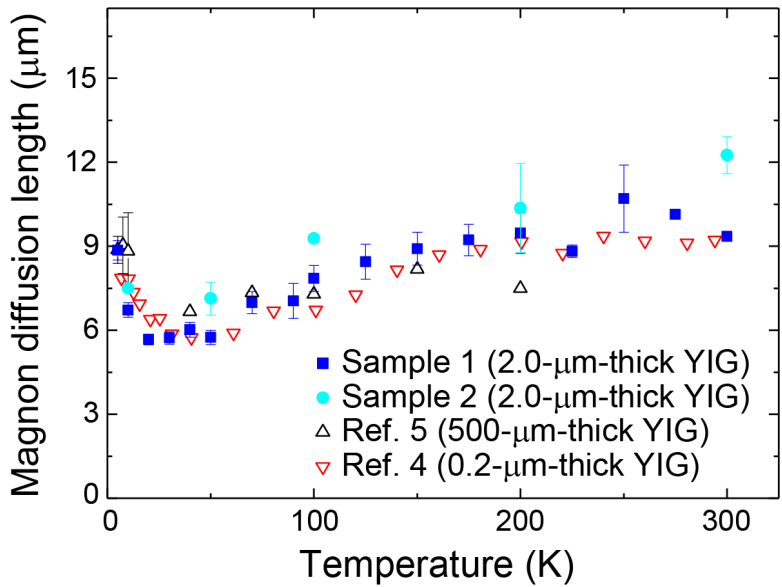

FIG. 5. Comparison of the temperature dependence of the magnon diffusion lengths of thermally driven magnon currents for different YIG thicknesses and growth methods: 2 μm (sample 1 and 2, this work; LPE), 500 μm (Ref. [5]; Czochralski method), and 0.2 μm (Ref. [4]; LPE).

**IV. Conclusions**

We performed nonlocal magnon transport experiments in 2-μm-thick YIG films and extracted the magnon diffusion lengths of electrically and thermally driven magnon currents in the linear regime as a function of temperature and magnetic field. The results reveal clear differences in the decay length of the magnon currents depending on the biasing method, with the thermally biased ones propagating to longer distances than those biased electrically. A significant, rather temperature-independent, suppression of the magnon diffusion length with applied magnetic field is observed for both biasing methods, pointing towards sub-thermal magnons to dominate the long-range transport. We attribute the dependence of the magnon diffusion length with the biasing mechanism to the generation of different non-equilibrium magnon distributions. We also show that, at low magnetic fields, the same temperature dependence and value of the magnon diffusion length is obtained for thermally biased magnons in YIG samples of different thicknesses and growth conditions, demonstrating the robustness of the measurement method and that this quantity is an intrinsic parameter of YIG. Our results thus call for more complex models to accurately describe diffusive magnon currents in magnetic systems. Experiments such as frequency-dependent coherent generation and propagation of magnon currents, or spatially-resolved Brillouin light scattering measurements [41] in YIG films of different thickness could shed some light on the origin of the differences in the magnon diffusion transport characteristics depending on the biasing method, magnetic field, temperature, and thickness of the magnetic layer.

**Acknowledgments**

The authors would like to acknowledge Camilo Ulloa, Joseph Barker, Koji Sato, Eiji Saitoh, and Rembert Duine for helpful discussions. The work was supported by the Spanish MINECO under the Maria de Maeztu Units of Excellence Programme (MDM-2016-0618), Project No. MAT2015-65159-R, and Project No. RTI2018-094861-B-100, and by the Regional Council of Gipuzkoa (Project No. 100/16). J.M.G.-P. thanks the Spanish MINECO for a Ph.D. fellowship (Grant No. BES-2016-077301).

**Appendix A: Current dependence of the nonlocal spin Seebeck effect signal at low temperatures.**

Figure 6(a) shows the angular-dependent nonlocal spin Seebeck signal $V_{NL}^{th}$, normalized to the square of the injected current, for different currents at the lowest temperature measured ($T$ = 2.5 K). The curves at lower currents do not overlap, evidencing a non-quadric dependence of the nonlocal voltage $V_{NL}^{th}$. This anomaly is also observed in Fig. 6(b), which shows that $V_{NL}^{th}$ does not scale with $I^2$. This anomalous dependence has already been reported in YIG at 3 K [31,32], but its origin remains unclear. Understanding this deviation from the expected behavior would require further studies that are beyond the scope of this work.

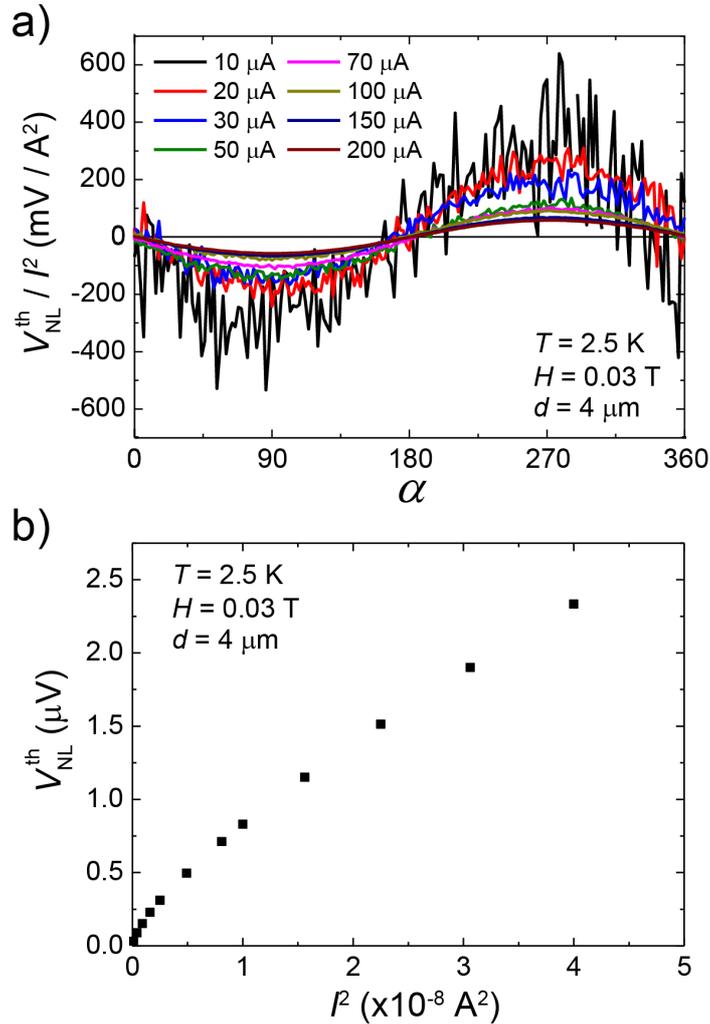

FIG. 6. (a) Angular dependence of the nonlocal signal $V_{NL}^{th}$ taken at 2.5 K for a magnetic field of 0.03 T rotating in the plane of the film and for different currents applied. The nonlocal voltage is normalized to the square of the applied current. (b) Amplitude of the nonlocal voltage extracted from data in panel a as a function of the square of the current. All data correspond to sample 1.

**Appendix B: Temperature dependence of the spin Hall magnetoresistance.**

The spin Hall magnetoresistance (SMR) is a local magnetoresistance that arises from the interplay between the spin accumulation generated in the Pt strip and the magnetic moments of the YIG layer [42,43] and is thus closely related to the electrical biasing of magnons studied here [2,3]. We evaluated the SMR in our devices by measuring the longitudinal resistance $R_L$ of the Pt strips while rotating the magnetic field in the plane of the film. The amplitude of the resistance modulation, $\Delta R_L$, is extracted from fitting the signal obtained to the expected $sin^2\alpha$ dependence [42–45], being the ratio $\Delta R_L/R_L$ the amplitude of the SMR. Figure 7 shows the temperature dependence of the SMR amplitude measured in different Pt strips of sample 1. The SMR amplitude is similar for the different Pt strips, demonstrating that the YIG/Pt interfaces have similar electron spin-magnetic moment coupling strength [46]. This result justifies the use of data points extracted from different devices varying the injector-detector distance $d$ to extract the magnon diffusion length.

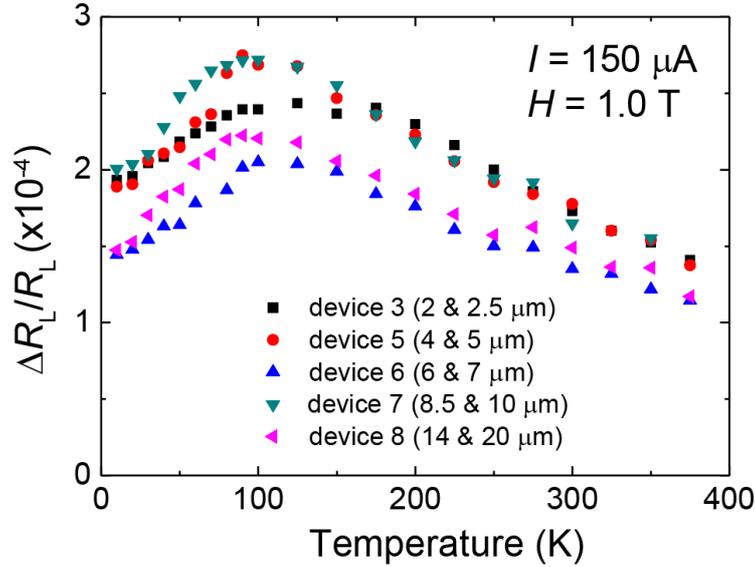

FIG. 7. Temperature dependence of the amplitude of the spin Hall magnetoresistance measured in different Pt strips at $H = 1.0$ T. A current of 150 μA was used. All data corresponds to Sample 1.

**Appendix C: One-dimensional and two-dimensional diffusion models to evaluate the magnon diffusion length.**

In previous reports, one- (1D) [3] or two-dimensional (2D) [47] models have been used to extract the magnon diffusion length, depending on the thickness of the MI. In our particular case of a 2-μm-thick YIG, we do not expect a 2D diffusion in the case of the thermally biased magnons because the fits are performed for injector-detector distances much longer (from 8.5 to 20 μm) than the thickness of the YIG film (2 μm). In addition, we fit the natural logarithm of the signal, which gives more weight to the long-distance measurements. Finally, following the recommendation of a previous study [10], distances in the range or longer than the magnon diffusion length are used. To properly evaluate the magnon diffusion length for thermally driven magnon currents, it is also important to avoid distances that are comparable to the YIG thickness because otherwise the YIG/GGG interface strongly influences the amplitude of the non-local signal [21]. Therefore, we are confident that the use of a 1D diffusion model for the fitting of the decay of thermally biased magnons is justified. On the other hand, for the case of the electrically driven magnon currents, the injector-detector distances investigated are shorter and, consequently, we have explored the possibility that the 2D model could better fit the magnon diffusion in our system.

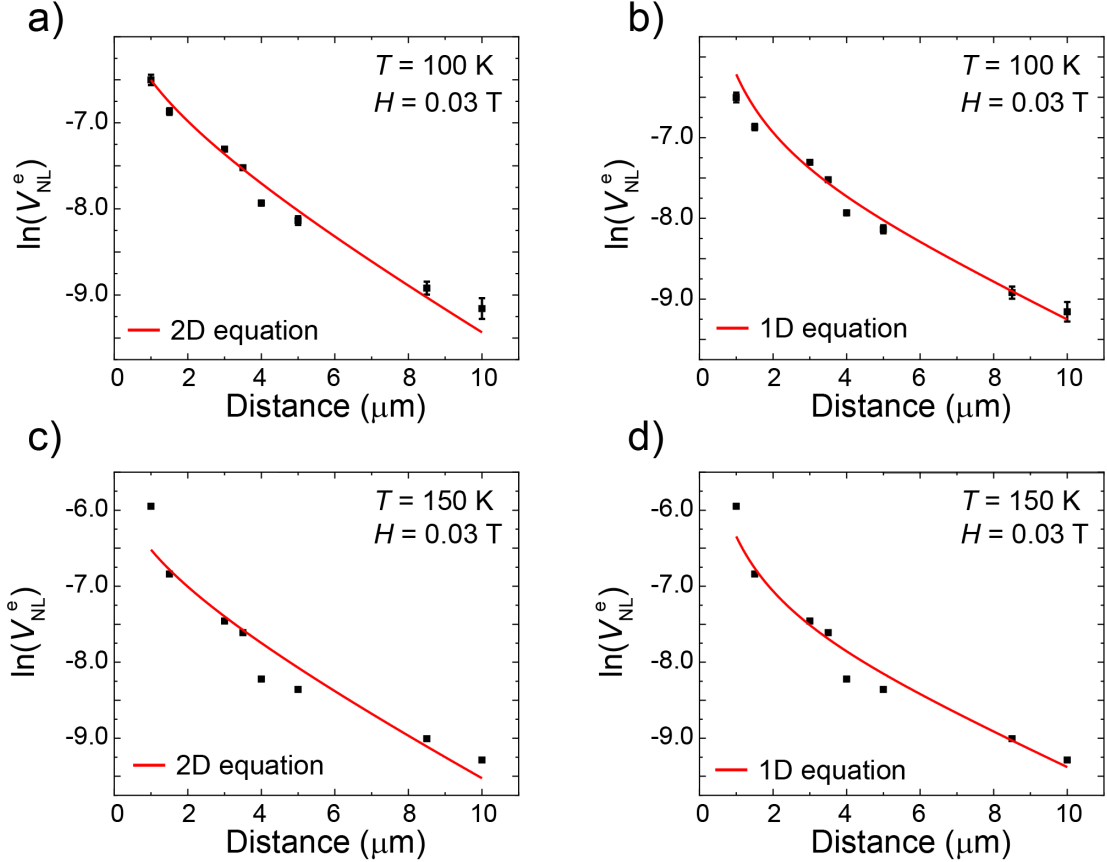

FIG. 8. Amplitude of the nonlocal voltage measured for electrically driven magnon currents at (a)-(b) 100 K and (c)-(d) 150 K. The measurement field is $H = 0.03$ T. The red lines are the fits to the (a),(c) 2D [47], and (b),(d) 1D diffusion models [3].

Figure 8 shows the fits to the nonlocal amplitude of the electrically biased magnons vs distance (at two representative temperatures, 100 K and 150 K, and $H = 0.03$ T) using the 1D diffusion model [3]:

$$V_{\text{NL}} = \frac{C}{\lambda_m} \frac{\exp(d/\lambda_m)}{1-\exp(2d/\lambda_m)} \quad (1)$$

and the 2D diffusion model [47]:

$$V_{\text{NL}} = A K_0\left(\frac{d}{\lambda_m}\right) \quad (2)$$

where $d$ is the distance between contacts, $\lambda_m$ is the magnon diffusion length, $K_0$ is the modified Bessel function of the second kind, and C and A are constants. As in the thermal case, the fits are performed to the natural logarithm of the nonlocal amplitude to give more weight to the long distances. Both models reproduce very well the experimental data, with values of the magnon diffusion length that are similar and compatible to the values presented in the main text (obtained by a single exponential fit in the relaxation regime), see Table 1. The fact that both models can fit the experimental data evidences that our system is an intermediate case between the 1D and the 2D model, as expected from the fact that the thickness of our YIG is on the same order, but lower, than the magnon diffusion length. Taking into account that the distances used for the experimental data are longer than both the thickness and the magnon diffusion length extracted, we kept the single exponential decay associated to the 1D model in the main text.

|  | $\lambda_m$ 1D model fit | $\lambda_m$ 2D model fit | $\lambda_m$ Single exponential fit |
|---|---|---|---|
| 100 K | 4.4±1.3 μm | 4.5±0.7 μm | 4.7±0.5 μm |
| 150 K | 4.4±0.9 μm | 4.4±0.9 μm | 5.4±0.5 μm |

Table 1. Magnon diffusion length values for electrically driven magnon currents extracted from the fits to the 1D and 2D model in Fig. 8, and to the single exponential for distances longer than $\lambda_m$ (as reported in the main text, see Fig. 3(a).

**Appendix D: Distance dependence of the nonlocal signal for thermally biased magnons.**

In order to determine the different regimes of the nonlocal thermal signals $V_{NL}^{th}$ in our YIG films, i.e., the exponential and $1/d^2$ regimes, to ensure that the magnon diffusion length is extracted from the data belonging to the exponential regime, we used sample 2 which contains devices with longer injector-detector distances than the devices in sample 1. Figure 9(a) shows the amplitude of the nonlocal signal for thermally biased magnons at $H = 0.03$ T as a function of $d$ and for different temperatures. Two different regimes are clearly identified: (i) for distances up to $d \sim 50$ μm, a first signal decay is identified, which corresponds to the expected exponential decay of the magnon chemical potential with distance [2–4,10,18,21]. The $\lambda_m^{th}$ extracted in sample 2 in this regime matches well with the values obtained in sample 1 [Fig. 5], confirming that the $\lambda_m^{th}$ extracted in both samples indeed corresponds to the magnon diffusion length of thermally excited magnons; (ii) for $d > 50$ μm, a second signal decay with an apparent longer characteristic length is identified. In this region, the system enters in the so-called $1/d^2$–regime, in which the signal is dominated by the temperature gradients at the YIG/GGG interface beneath the Pt detector [10]. This result is indeed expected for our YIG thickness and $\lambda_m^{th}$ values, in agreement with Ref. [10]. The two regimes can be clearly distinguished for $H = 0.03$ T and 2 T [green open and black solid squares in Fig. 9(b), respectively]. However, at $H = 5$ T and 9 T [red solid circles and blue solid triangles and green squares in Fig. 9(b), respectively], it is difficult to evaluate the existence of both regimes as no clear change in the slope of the signal decay with $d$ can be identified, indicating that, at the largest measured distances, both magnon transport and nonlocal thermal gradients might contribute to the nonlocal $V_{NL}^{th}$ signal. Note that, with increasing the magnetic field, $\lambda_m^{th}$ decreases and then the emergence of the $1/d^2$ regime would shift to shorter distances. In this case, and only for these large magnetic fields, we could be overestimating $\lambda_m^{th}$. The saturation of $\lambda_m^{th}$ with magnetic field above ~2 T might be indicating that this is the case [Fig. 3(f)].

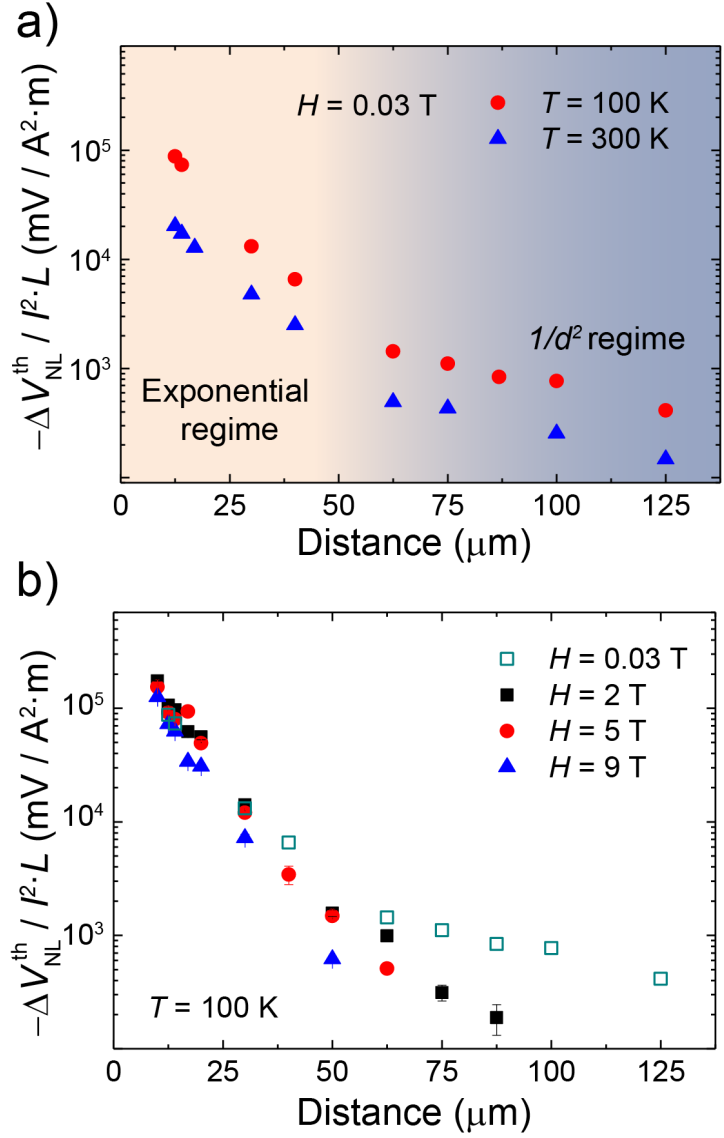

FIG. 9. (a) Distance dependence of the amplitude of the nonlocal voltage for thermally driven magnon currents taken at different temperatures in Sample 2. The signal is normalized to the square applied charge current (150 μA) and the length of the detector ($L$). The measurement field is $H = 0.03$ T. Two different regimes are identified: the exponential regime, for $d < 50$ μm, and the $1/d^2$–regime, for $d > 50$ μm. (b) Distance dependence of the nonlocal voltage for thermally driven magnon currents at $T = 100$ K and different magnetic fields. The two regimes are clearly distinguishable for low magnetic fields ($H = 0.03$ T and 2 T). However, at high magnetic fields, it is difficult to evaluate the distance at which the threshold between the two regimes occurs.